\documentclass[twocolumn,showpacs,preprintnumbers,amsmath,amssymb,aps]{revtex4}

\usepackage{graphicx}
\usepackage{dcolumn}
\usepackage{bm}
\usepackage{bbm}

\begin{document}

\title{Quantum Filtering of Optical Coherent States}

\author{C. Wittmann$^1$}
	\email{cwittmann@optik.uni-erlangen.de}
\author{D. Elser$^1$}
\author{U.L. Andersen$^{1,2}$}
\author{R. Filip$^{1,3}$}
\author{P. Marek$^{4}$}
\author{G. Leuchs$^1$}
\affiliation{
$^1$Institut f\"ur Optik, Information und Photonik, Max-Planck Forschungsgruppe, Universit\"at Erlangen-N\"urnberg,
G\"unther-Scharowsky-Stra{\ss}e 1, 91058, Erlangen, Germany\\
$^2$Department of Physics, The Technical University of Denmark, 2800 Kongens Lyngby, Denmark\\
$^3$Department of Optics, Palack\'y University,
17. listopadu 50, 772 07 Olomouc,
Czech Republic\\
$^4$School of Mathematics and Physics, Queen's University, Belfast BT7 1NN, United Kingdom
}

\date{\today}

\begin{abstract}
We propose and experimentally demonstrate non-destructive and noiseless removal (filtering) of vacuum states from an arbitrary set of coherent states of continuous variable systems. Errors {i.e.} vacuum states in the quantum information are diagnosed through a weak measurement, and on that basis, probabilistically filtered out. We consider three different filters based on on/off detection phase stabilized and phase randomized homodyne detection. We find that on/off detection, optimal in the ideal theoretical setting, is superior to the homodyne strategy in a practical setting.
\end{abstract}

\pacs{03.67.-a, 03.67.Hk}

\maketitle

\section{Introduction}

Ultra-low noise quantum channels transmitting discrete or Continuous-Variable (CV) quantum information are prerequisite for the successful execution of many quantum information protocols. For example, the security and the secret key rate of quantum key distribution critically depend on the amount of excess noise added to the quantum state during transmission~\cite{Grosshans.Nature421,Heid.PRA73}. All realistic quantum channels are afflicted by such noise:
In fiber channels, for example, light scattering by thermal phonons causes Gaussian phase noise. On the other hand, noise sources important in atmospheric transmission, such as time jitter and beam pointing noise~\cite{Gisin.RMP74}, show a characteristic non-Gaussian behavior.

In order to retain security, the errors imposed by the noisy channels must be corrected. Various methods have been developed to combat noise in CV quantum communication, examples being entanglement distillation \cite{Duan.PRL84} and quantum error correction coding \cite{Braunstein.Nature394}, which are relying on highly non-classical resources and complex processing.
An alternative is quantum filtering which is a protocol that probabilistically rejects erroneous quantum states through detection. The simplest approach is a classical
measure-prepare strategy based on optimal state
discrimination using the Neyman-Pearson criterion
\cite{Neyman.PTRSL231} followed by state recreation.
Helstrom found
that by using a tailored detection process, it is possible
to identify a pure target state in a noisy mixture
\cite{Helstrom.book} (see also~\cite{Paris.PRA225,D'Ariano.PRA65,Takeoka.PLA313}).
Takeoka et al. generalized this strategy and named it unambiguous quantum state filtering since it unambiguously
filters out a specific signal from the noise~\cite{Takeoka.PRA68}. However, only a single a priori known state is resurrected, which is done destructively and therefore not suitable for quantum communication.

In this paper, we propose and experimentally realize a quantum state filter protocol specially tailored to non-Gaussian noise as in atmospheric transmission. The protocol filters a coherent state alphabet non-destructively and noiselessly, {i.e.} the quantum states are not completely destructed and no excess noise is added by our filter. Our protocol is based on a weak measurement of the corrupted signal followed by a post selection of the remaining part of the signal. We investigate two different weak measurement strategies, namely homodyne detection and on/off detection and compare their efficiencies in filtering out noise.  We find that optimum filtering is obtained by the use of an ideal on/off detector.
The scheme presented in this paper provides the first implementation of a CV error detection protocol enabled by a photon counting detector.

\begin{figure} [h]
\centerline{\includegraphics[width=0.45\textwidth]{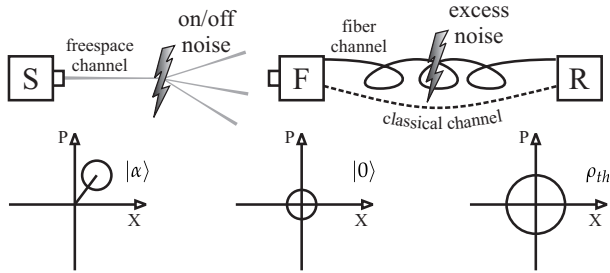}}
\caption{\label{fig:explanation} Application of the quantum filter device. The filter F is placed between two quantum channels connecting sender S and receiver R. We assume, that the channels have non-Gaussian on/off (first part) and Gaussian properties (last part). The on/off behavior of the first channel will be masked by excess noise in the second channel (e.g. $|\alpha\rangle\langle\alpha|\rightarrow|0\rangle\langle0|\rightarrow\rho_{th}$). However, a quantum filter in the intermediate station can sense the channel break and reject the noisy state by sending information over a classical channel to R.}
\end{figure}

An exemplary application of such a filter is shown in Fig.~\ref{fig:explanation}. Suppose a signal is conveyed through two different quantum channels each possessing different kinds of noises (e.g. a free space channel and a fiber channel). If the first channel is inflicted by the non-Gaussian on/off noise and the following channel by Gaussian noise, the on/off noise might be completely masked by the Gaussian noise and cannot easily be filtered out at the receiving station. In order to circumvent a mixing of the two noise sources, the filtration station could be placed between the two channels thus removing the on/off noise before the signal enters the Gaussian noise channel.

The filtration protocol can be also used to improve the security of a quantum key distribution scheme based on a coherent state alphabet and heterodyne detection. This is proven at the end of the paper.

\section{Description of the protocol}

Let us consider the protocol in detail. Information is encoded into quantum states taken from a coherent state alphabet with a possibly unknown probability distribution.
The quantum state is subsequently sent through the quantum channel where it is subject to time jitter or beam positioning noise. Such non-Gaussian noise occurs when the detection time is longer than the signal but shorter than the jitter time or when the aperture of the receiver is larger than the beam but much smaller than the beam pointing noise. This noise can be approximated by a mixture of the sent coherent state $|\alpha\rangle$ and the vacuum state:
\begin{equation}
\rho(\alpha)=p|\alpha\rangle\langle\alpha|+(1-p)|0\rangle\langle 0|,
\end{equation}
where $p$ is an unknown probability for perfect transmission. The task is now to find a protocol that unambiguously filters out the vacuum state, while only attenuating the coherent state, e.g. $|\alpha\rangle\rightarrow|\gamma\alpha\rangle$, $\gamma<1$.

To accomplish a state independent weak measurement adding no excess noise the signal system must be coupled unitarily and phase insensitively to a meter system in which the actual measurement takes place. Due to these requirements, the coupling can be  enabled by a beam splitter with the meter system being in the vacuum state before interaction~\cite{Leonhardt.book}. The signal-meter coupling can therefore simply be described by the transformation
\begin{equation}
\rho (\alpha )\otimes\rho (0)\rightarrow \rho(\sqrt{1-R}\alpha)\otimes\rho(\sqrt{R}\alpha),
\end{equation}
where $R$ is the reflectivity of the beam splitter.
After this interaction, the presence or absence of the vacuum contribution is correlated in the two systems. Thus by detecting the vacuum state in the meter system, filtering of the vacuum in the signal system can be performed by post selecting on the correlated state. The strategy is illustrated in Fig.~\ref{fig:explanation}. The next step is thus to find the measurement strategy that optimally and unambiguously detects the vacuum contribution.

Let us assume that we use the measurement operators $\Pi^\bot$ and $\Pi$ to discriminate the vacuum state and the unknown signal state. We seek a strategy that maximizes the probability $\langle\sqrt{R}\alpha | \Pi|\sqrt{R}\alpha \rangle$ of measuring $ |\sqrt{R}\alpha \rangle$ under the condition that the vacuum state is never detected incorrectly, that is $E=\langle 0|\Pi|0\rangle=0$.
Such decision problem was first encountered by Neyman and Pearson~\cite{Neyman.PTRSL231} and was further elaborated upon by Helstrom~\cite{Helstrom.book} and Holevo~\cite{Holevo.book}. They found that the maximum probability of detecting the signal correctly (also called the acceptance probability) with no error detections ($E=0$) is given by $P(\sqrt R \alpha)=1-\mathrm{exp}(-R|\alpha|^2)$ (Note that $E=P(0)$). We readily find that measurement operators satisfying these conditions are $\Pi^\bot=|0\rangle\langle 0|$ and $\Pi=\mathbbm{1}-|0\rangle\langle 0|$ for rejecting and accepting the state, respectively. Therefore, using these measurement operators the signal states can be unambiguously detected in the meter system and thus perfectly filtered out in the signal system. We stress that since this optimized measurement is independent of the signal amplitude and the reflection coefficient, it is the optimal strategy for every coherent state.

The physical implementation of these measurement operators is known to be an ideal avalanche photodiode (APD) operating in the break down voltage mode. Practical APDs are, however, lossy and possess dark counts which results in a reduced success probability and gives rise to errors, that is, $E>0$. Therefore, in addition to the APD, we investigate in the following the filtering performance using a homodyne detector for the decision problem; quadrature values larger than a certain a priori specified threshold value are assumed to stem from the unknown signal state, smaller values from the vacuum state. Note that a similar strategy was proposed in Ref.~\cite{Suzuki.PRA73} and experimentally realized in Ref.~\cite{Heersink.PRL96,Franzen.PRL97} to purify non-classical resources. We also note that the incorporation of an APD in a CV system has been implemented in previous experiments on state preparation and estimation~\cite{Wenger.PRL92,Wenger.PRA70}.

\begin{figure}
\begin{tabular}{l}
(a) \\ [-0.3cm]
            \centerline{\includegraphics[width=0.35\textwidth]{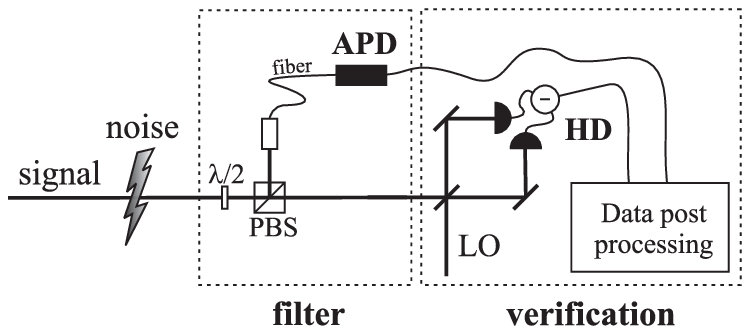}} \\ [+0.1cm]
        (b) \\ [-0.1cm]
        \centerline{\includegraphics[width=0.47\textwidth]{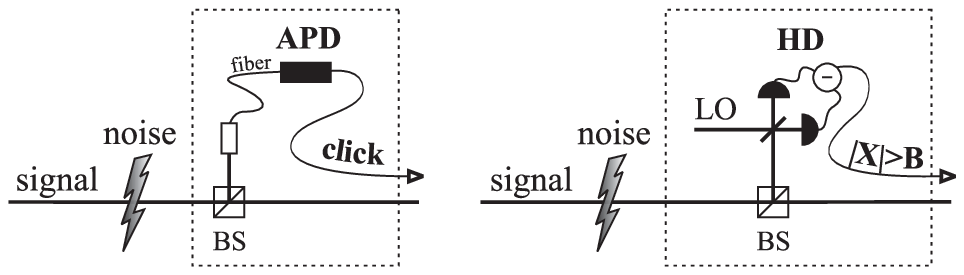}} \\
        \end{tabular}
    \caption{\label{fig:Setup} Schematic illustration of a coherent state quantum filter for the non-Gaussian channel: (a) Filter device with verification measurement; (b left) Filter using APD as a detector, (b right) using homodyne detection with a local oscillator (LO). }
\end{figure}

\begin{figure}
        \includegraphics[width=0.45\textwidth]{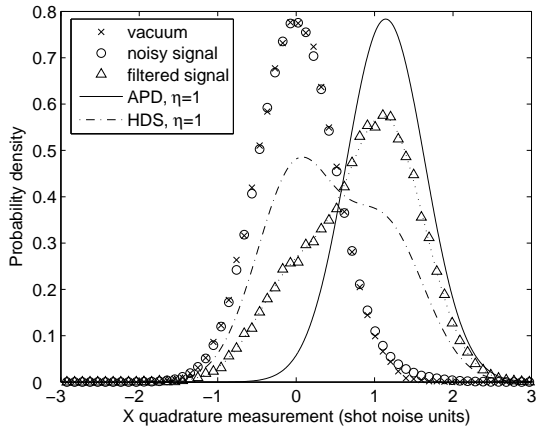}
    \caption{\label{fig:etha0,01_PX_X} Marginal distribution for the perturbed state ($p=0.02$) (circles), the vacuum state (crosses) and the filtered state using an APD filter (triangles).
    The solid and the dotted dashed line correspond to the theoretical performance of a filter with APD ($\eta_{\rm{APD}}=1$) and with homodyne detector ($\eta_{\rm{HDS}}=1$) respectively. The mean photon number in the filter is $R|\alpha|^2=1.65$ and the error probabilities are identical $E_{\rm{APD}}=E_{\rm{HDS}}=5.3\cdot10^{-3}$.}
\end{figure}

\begin{figure}
        \includegraphics[width=0.45\textwidth]{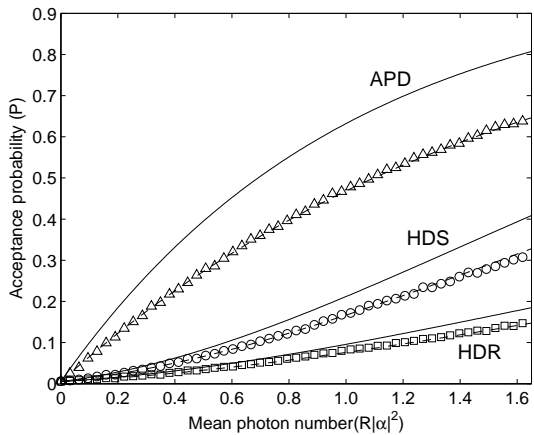}
    \caption{\label{fig:etha0,02_P_R} Acceptance probability for mean photon number $R|\alpha|^2$ impinging on the filter detector. The  triangles, circles and squares show experimental data for APD and homodyne detection with and without stabilized LO, respectively. The solid lines are theoretical predictions for detectors with unit quantum efficiency. $E_{\rm{APD}}=E_{\rm{HDS}}=E_{\rm{HDR}}=5.3\cdot10^{-3}$. }
\end{figure}

In order to quantify the performance of the filtering protocol using different detection methods, we introduce two appropriate functions: the sensitivity $S$ and the gain $G$. The sensitivity quantifies the filtering efficiency near the vacuum state and we define it as
\begin{equation}\label{Sensitivity}
S=\frac{1}{2}\frac{\mathrm{d}^2}{\mathrm{d}|\alpha|^2}P(\sqrt R \alpha)|_{\alpha=0}.
\end{equation}
Since the probablity $P$ must be minimal when $\alpha =0$, the sensitivity $S$ is a measure for how quickly the probability increases around $\alpha =0$. For the ideal filter we easily find $S=R$, thus we will be using $S/R$ as the figure of merit. The other parameter that we will use to quantify the performance of the filter is the gain $G=p'/p$ where $p'$ is the probability for the coherent state to occur in the mixture after filtering: $\rho' =p'|\sqrt{T}\alpha\rangle\langle\sqrt{T}\alpha| +(1-p')|0\rangle\langle 0|$. The success probability for positive filter outputs is $P_S=pP(\sqrt R \alpha)+(1-p)P(0)$ and the gain can thus be written as
\begin{equation}\label{Gain}
G=\frac{1}{p}\left(1-(1-p)\frac{E}{P_S}\right).
\end{equation}
Note that the sensitivity $S$ depends solely on the filter implementation. Thus, it is independent of the channel. In contrast, the gain is a signal-, channel- and filter-dependent parameter, and therefore describes the joint action of channel and filter.

\section{Experimental setup and results}

\begin{figure}
\begin{tabular}{l}
\begin{minipage}{7.2cm}
\includegraphics[width=1.0\textwidth]{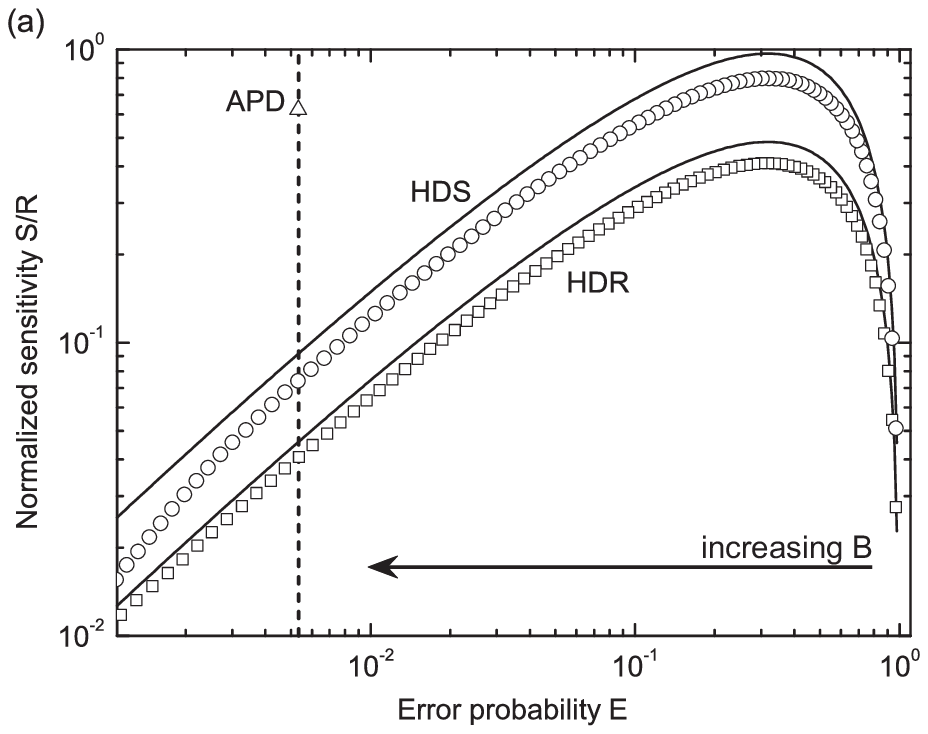}
\end{minipage}
\\[0.4cm]
\begin{minipage}{7.2cm}
\includegraphics[width=1.0\textwidth]{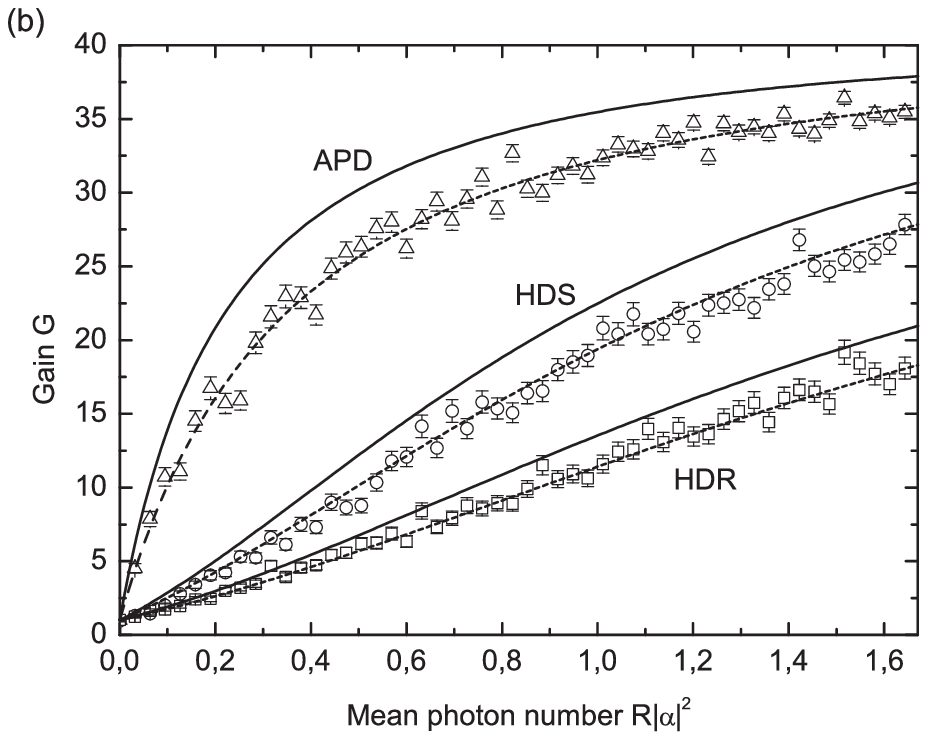}
\end{minipage}
\\[0.4cm]
\begin{minipage}{7.2cm}
\includegraphics[width=1.0\textwidth]{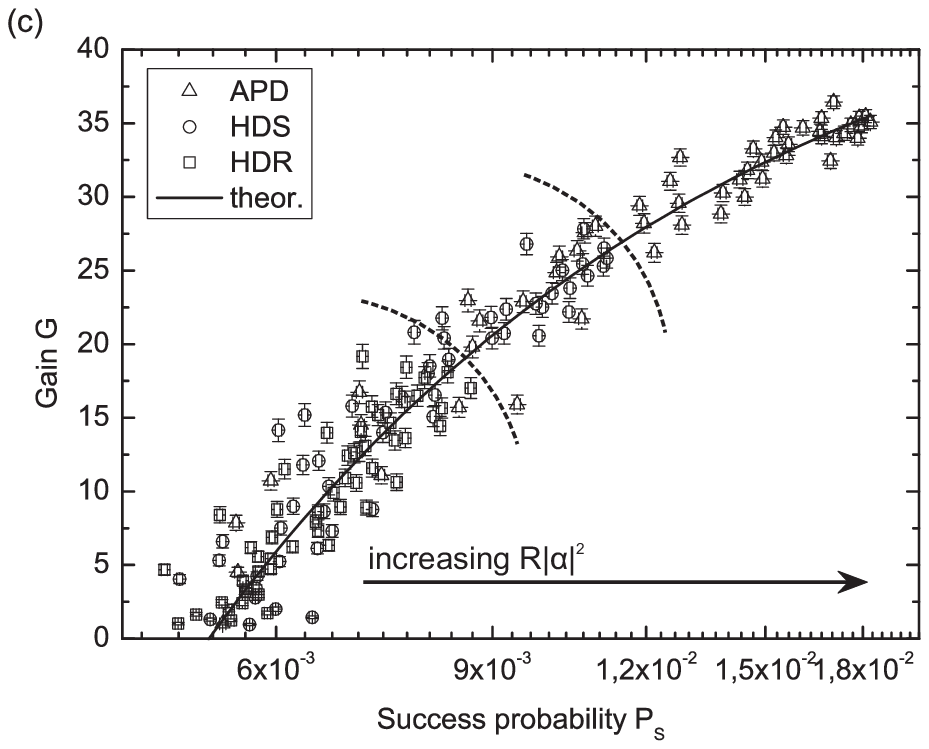}
\end{minipage}
\end{tabular}
\caption{\label{fig:etha0,02} In all figures the triangles are experimental data points for a filter using APD, the circles and squares show data for filters using homodyne detection with and without stabilized LO respectively; the solid lines are theoretical predictions for filters using unit quantum efficiency detectors, $\eta=1$ (APD: $S/R=(1-p_d)^2$ for $\eta=1$). (a) Sensitivity as a function of error probability. The dashed line should guide the eye to the error rate where the detectors are compared $E_{\rm{APD}}=E_{\rm{HDS}}=E_{\rm{HDR}}=5.3\cdot10^{-3}$. (b) Gain $G$ as a function of mean photon number $R|\alpha|^2$ impinging on the filter detector. Signal probability fixed to $p=2 \%$. (c) Parametric plot of Gain $G$ and success probability $P_S$ for $R|\alpha|^2\in[0,1.65]$. The performance decreases from APD to HDS and HDR. The error bars show the statistical errors (three-sigma error bars) and seem to be much smaller than the experimental errors.}
\end{figure}

We proceed with the experimental demonstration of quantum filtering using an APD-based filter. The setup is shown in Fig.~\ref{fig:Setup}(a). The source is a diode laser emitting light at $810\rm{nm}$, characterized with a coherence time of $1\rm{\mu} s$ and measured to be shot noise limited in the detected bandwidth. The statistical mixture of the coherent signal states and vacuum states is prepared in a computer controlled electro-optical modulator (EOM). The EOM therefore generates the signal and simulates the noisy channel. 
To obtain a small excitation of the coherent state, the beam is heavily attenuated after modulation. We set the coherent state probability to $p=2\%$, thus the probability for vacuum to occur being $1-p=98\%$. The signal duration is defined within 800ns time windows, while the rate of the signal preparation is set to 100kHz. 

We investigate the performance of the APD-based filter by characterizing the state with homodyne detection before and after filtering. First we demonstrate the principles of the protocol using the APD as a filter and the homodyne detector for characterization. The probability distribution of the mixed input state is shown in Fig.~\ref{fig:etha0,01_PX_X} by circles and for comparison the pure vacuum state is shown by crosses. Next, the mixed state is passing the filter beam splitter and the homodyne quadrature data are selected based on the measurement outcomes of the APD detector. The resulting probability distribution is shown by the dotted curve and should be compared with the ideally filtered signal (solid line) and two other curves: the one expected using homodyne detection with equal error probability and one using unit quantum efficiency as filter (dotted dashed line).

In the following, we fully characterize the quantum filters. The mixed state is split on a 50/50 beam splitter and subsequently directed to two different detector units: a fiber coupled APD (Perkin-Elmer SPCM CD3017) and a homodyne detector. We therefore simultaneously measure the acceptance probabilities for the two different detection schemes; here each detector represented a filter.

Let us first discuss the APD-based filter (see Fig.~\ref{fig:Setup}(b left)). A gate option is used to precisely determine the detection time. The quantum efficiency is estimated to be $\eta_{\rm{APD}}=63\pm3\%$, while the dark count rate is 180cts/s. Due to these imperfections, the expected acceptance probability is
\begin{equation}\label{ProbOK-Apd}
    P_{\rm{APD}}(\beta) = 1 -
    (1-p_d)\exp\left(-\eta_{\rm{APD}}(1-p_d)|\beta|^2\right)
\end{equation}
where $\beta=\sqrt{R}\alpha$ and $p_d$ is the dark count probability. The expected error probability is $E_{\rm{APD}}=P_{\rm{APD}}(0)=p_d$. We measure the acceptance probability by comparing the actual decision (based on the filter measurement outcome) with the a priori known preparation of the state. The results are presented in Fig.~\ref{fig:etha0,02_P_R} as a function of $R|\alpha|^2$. Note that $R$ should be tailored to the actual amplitude $\alpha$ to optimize the performance. The error probability is found to be $E=5.3 \cdot 10^{-3}$, which is limited not by the dark count probability $1.4\cdot 10^{-4}$ but by the imperfections in preparing the vacuum state.

Next we discuss the filter based on homodyne detection (see Fig.~\ref{fig:Setup}(b right)). The detector's bandwidth is 10MHz. Using a local oscillator (LO) power of about 5mW, the shot noise to electronic noise ratio is 18dB. The detection efficiency, including the mode matching efficiency and the quantum efficiency of the photodiodes is $\eta_{\rm{HD}}=84\pm3\%$. We investigate two different kinds of homodyne detectors: one with a phase stabilized local oscillator (HDS) and one with a phase randomized local oscillator (HDR). The latter scheme should be used when the input alphabet of coherent states is rotationally symmetric in phase space whereas the former scheme is superior if e.g. a binary phase encoding is used where the absolute direction of the displacement is known a priori. The hypothesis whether a signal or a vacuum state was measured is based on the absolute value of the measured quadrature; if it is above a certain threshold value, denoted $B$, we estimate the state to be $|\alpha\rangle$, if not $|0\rangle$. Knowing that the signal is encoded into $|\alpha\rangle$, the expected acceptance probability for a phase stabilized local oscillator  is
\begin{equation}\label{ProbOK-HDXP}
P_{\rm{HDS}}(\beta) =
\frac{\mbox{Erfc}[\sqrt{2}(B+a)]+\mbox{Erfc}[\sqrt{2}(B-a)]}{2}
\end{equation}
where $a=\eta_{\rm{HD}}\beta$, while for the phase randomized local oscillator
\begin{eqnarray}\label{ProbOK-HDX}
P_{\rm{HDR}}(\beta)
&=&\frac{1}{2\pi}\int_{-\pi}^{\pi}\mbox{Erfc}\left[\sqrt{2}(B-a\cos\theta)\right]d\theta.
\end{eqnarray}
The error is identical for the two approaches and given by
$E_{\rm{HDR}}=E_{\rm{HDS}}=\mbox{Erfc}[\sqrt{2}B]$.

A classical signal is appended to the pulse trains to estimate the phase difference of signal and LO at a given time. Measurement data, acquired with randomized phase of the LO, is subsequently directly used to evaluate the performance of the scheme based on random LO. However, due to the appended classical signal the relative phase is known and we selected the data associated with a phase difference of zero corresponding to the phase stabilized case. Using these data and the above hypothesis, we find the acceptance probability for various excitations $R|\alpha|^2$ and various threshold values B. In Fig.~\ref{fig:etha0,02_P_R} we plot the acceptance probability as a function of $R|\alpha|^2$ with the threshold value set such that the error probability matches the one of the APD. It is clear from the plot that the APD performs better than the homodyne detector despite the much higher quantum efficiency of the latter one.

The sensitivity of the two filters is obtained by fitting curves to the measured acceptance probabilities corresponding to various thresholds (for the homodyne case) and subsequently using equation~(\ref{Sensitivity}). The results are plotted in Fig.~\ref{fig:etha0,02}(a), where the triangle represents the APD-based filter whereas the circles and  squares are associated with the phase randomized and phase locked LO, respectively.  For post selection thresholds B close to the coherent state variance the sensitivity of homodyne detectors is maximal. As evident from the plot, for identical error probability the sensitivity of the APD-based filter is much larger than that of the homodyne-based filter.

From the measurements we also calculate the gain for different mean photon numbers and different success probabilities as shown in Fig.~\ref{fig:etha0,02}(b) and \ref{fig:etha0,02}(c), respectively. The former figure clearly shows the superior performance of the APD filter compared with the homodyne filters. From Fig.~\ref{fig:etha0,02}(c) we clearly see that the behavior of the gain as a function of the success probability follows the same curve for the three filters if the error rates are tailored to be the same (by adjusting the error thresholds appropriately). This is also what is expected from Eq.~(\ref{Gain}).

\section{Filtering in a quantum key distribution scheme}
\label{FilterInQKD}

In the final part, we consider the filtering action in a CV quantum key distribution scheme using Gaussian modulated coherent states~\cite{Grosshans.Nature421}. To estimate a lower bound for secure transmission we use the recent work of Garcia-Patron and Cerf~\cite{Garcia.PRL97} (see also ref~\cite{Navascues.PRL06}), showing that the lower bound can be directly computed from the covariance matrix of the joint state between the sender of information, Alice, and the receiver, Bob. The lower bound (for reverse reconciliation) is thus given by
\begin{equation}\label{lower}
K_{lower}=I_{ab}^G-\chi_{bE}^G
\end{equation}
where $\chi_{bE}^G$ is the maximum information between Bob and the eavesdropper, Eve, corresponding to the Holevo bound and $I_{ab}^G$ is the Gaussian mutual information between Alice and Bob. These quantities are computed solely from the covariance matrix of the joint state of Alice and Bob.

The joint state is found by using the equivalence between the coherent state scheme and an entanglement-based protocol. In an ideal entanglement based protocol, Alice generates Gaussian two-mode squeezed states and measures one mode using heterodyne detection. The other mode is then being prepared in a coherent state with a Gaussian distributed displacement. A two-mode squeezing variance of $V$ results in a displacement variance of $\sigma=(V+1/V)/2-1$. The prepared coherent state is sent to Bob through the erasure channel and produces the following density matrix
\begin{equation}\label{jitter}
\rho_{AB}=p\left(|V\rangle\langle V|\right)_{AB}+(1-p)\rho_A\otimes \left(|0\rangle\langle 0|\right)_{B},
\end{equation}
where $|V\rangle_{AB}$ represents a two-mode squeezed state and $\rho_A$ is a thermal state having a variance $(V+1/V)/2$. The lower bound for secure communication is found numerically by solving $K_{lower}<0$ using the covariance matrix of the joint state $\rho_{AB}$ and Eqn.~(\ref{lower}). We find security cannot be guaranteed if $p<0.87$ (corresponding to $K_{lower}<0$). $K_{lower}$ is maximized over the variance $V$.

Let us now consider the security when the filtering protocol is implemented. Assuming first the APD to be ideal, the state after filtering is
\begin{equation}\label{jitter}
\rho_{filter}=\frac{\Pi U_{BS}\rho_{ABT} U_{BS}^\dagger\Pi}{\mbox{Tr}(\Pi U_{BS}\rho_{ABT} U_{BS}^\dagger)}
\end{equation}
where $U_{BS}$ is the unitary beam splitter operation, $\rho_{ABT}=\rho_{AB}\otimes |0\rangle\langle 0|$ and $\Pi$ is the measurement operator of the tap T.
The structure of the covariance matrix for this state is given by
$CV'_{AB}=\frac{1}{P_S}(CV_{AB}-P^0 CV^0_{AB})$,
where $CV_{AB}$ is the covariance matrix of the state right after the beam splitter after tracing out the tap mode, $\rho_{BS}=\mbox{Tr}_{T}(U_{BS}\rho_{ABT} U_{BS}^\dagger)$, and $CV^0_{AB}$ is the covariance matrix of the state if the measurement outcome of the filter measurement is associated with $\Pi^{\bot}=1-\Pi$, $\rho^\bot=\Pi^{\bot} U_{BS}\rho_{ABT} U_{BS}^\dagger\Pi^{\bot}/\mbox{Tr}(\Pi^{\bot} U_{BS}\rho_{ABT} U_{BS}^\dagger)$. $P^0$ is the probability for
getting a measurement result associated with $\Pi^{\bot}$ and $P_S=1-P^0$ is the filtering success rate.
Using the covariance matrix of the filtered state, $CV'_{AB}$, we again compute the lower bound numerically, and find that the probability $p$ for which secure communication can take place is now just required to be larger than zero. This can be also investigated analytically in the limit of weak two-mode squeezing ($V\approx 1$), corresponding to a small Gaussian alphabet. In this case, the lower bound on the secure key rate is approximately
\begin{equation}
K_{lower}\approx pP_S\frac{1}{2}\log_2\left[\frac{\mathrm{e}}{2} \right]T(V-1)^2,
\end{equation}
which is positive for any $T>0$ and $V\not=1$. We have thus shown that the usages of an ideal filter reestablishes the security of the quantum key distribution system independent on the amount of vacuum noise.


We now consider the realistic case where the APD is nonideal. Using an APD with nonzero quantum efficiency is not a hindrance for obtaining secure communication, but it will result in a lower success rate. A limiting factor on the security, however, is the presence of dark counts which limits the minimum noise probability $p_{min}$ for which security can be proven.
For a given dark count rate, the transmission $T$ of the filtering beam splitter and the modulation variance $\sigma$ can be numerically optimized to maximize $K_{lower}$ for which security can be guaranteed in the protocol with respect to the noisy channel.
Using the experimental parameters of the APD used in our experiment ($\eta=0.63$, $p_d=0.005$) we find that secure communication can be guaranteed if $p>p_{\mathrm{min}}=0.222$. By reducing the dark counts to $p_d=5\times 10^{-4}$ the minimum probability is reduced to $p_{\mathrm{min}}=0.028$ and for even lower $p_d=5\times 10^{-5}$, the threshold is only $p_{\mathrm{min}}=0.003$.


\section{Conclusion}

In summary, we have investigated a filtering protocol that successfully filters out vacuum states from a set of coherent states of a continuous variable system. A weak measurement scheme consisting of a beam splitter and an optimized measurement was employed to probabilistically filter out the unwanted vacuum states. Different measurement strategies based on homodyne and on/off detection were investigated and compared. We therefore provided the first direct comparison between an APD-based and a homodyne-based protocol, and found that the ideal on/off detection is optimal and that the practical (that is non-ideal) on/off detector is superior to homodyne detection despite the much higher quantum efficiency of the latter one. The protocol will be advantageous in continuous variable quantum communication wherever beam positioning and time jitter noise are the main obstacles for faithful and secure transmission.

\medskip
\noindent {\bf Acknowledgments}

This work was supported by the EU project COVAQIAL (project no. FP6-511004), the EU-IST network SECOQC, COMPAS project no. 212008, the project 202/07/J040 of the GACR and project MSM6198959213 of Czech Ministry of
Education. R.F. acknowledges
support from the Alexander von Humboldt foundation. P.M. acknowledges
support from the European Social Fund.

\bibliographystyle{apsrev}

\end{document}